\documentclass{JAC2003}
\usepackage{graphicx}
\usepackage{axodraw}
\begin{document}
\title{Non $q\overline{q}$ light meson spectroscopy.}

\author{Felipe J. Llanes-Estrada \thanks{fllanes@fis.ucm.es \ \ 
(on leave at Inst. f\"ur Theoretische Physik, Auf der Morgenstelle
14, D-72076 T\"ubingen, Germany).}, \\
Depto. F\'{\i}sica Te\'orica I, Univ. Complutense de Madrid, 28040 Madrid 
Spain \\}

\maketitle

\begin{abstract} 
In this talk I comment on some theoretical expectations
for exotic light meson spectroscopy below 2 GeV and their potential 
interest for a future energy upgrade of DAFNE.
\end{abstract}

\section{Matters of principle}

The colours we perceive around us vary almost continuously between
different tones. Only careful scrutiny of the light emitted by pure 
substances through diffraction gratings at the end of the 19th
century demonstrated the  separation of these colours into discrete lines, 
opening a window to a new world of phenomena.
In the same way we hope that the spectral lines that form the light of 
the strong interactions will be resolved and the energy differences 
between them will help us understand the dynamics of the strong force in 
detail.
With this statement of purpose in mind, in this note I comment on the 
spectroscopy of light mesons, with special 
attention to non conventional $q\overline{q}$ states.

\subsection{The mass gap.}
The first observation I would like to make is that the theory of the 
strong interactions, Quantum Chromodynamics (QCD) {\it must} present
a mass gap. To understand it, think of the meson Fock space of 
this quantum field theory, 
$$
\arrowvert \overline{q} q \rangle ,  \ \ 
\arrowvert \overline{q} q \overline{q} q \rangle , \ \
\arrowvert \overline{q} q g \rangle , \ \ 
\arrowvert \overline{q} q g \rangle , \ ...
$$
given a state with any particular quantum numbers and a given mass,
we could construct another state with the same quantum numbers and
extremely close mass by just adding a pair of the current partons to it.
Therefore there would be no discrete part to the spectrum at all. This
may in fact not be such a bad approximation at high masses, but the low 
lying states follow definitely a discrete pattern. Therefore, adding
a quark-antiquark pair or a couple of gluons to any one state must have
an energetic cost, the mass gap. The constituent quark model tells us
what the cost of a constituent light quark (u or d) is: $300 \ MeV$, about
a third of the mass of the proton. For the strange quark $500 \ MeV$ 
(half of the mass of the $\phi$ meson) is just about right. There is no 
consensus on the gluon mass gap, I will give below   
some indications on this respect.

\subsection{How many mesons do we know of?}
\begin{figure}[htb]
\centering
\includegraphics*[width=65mm]{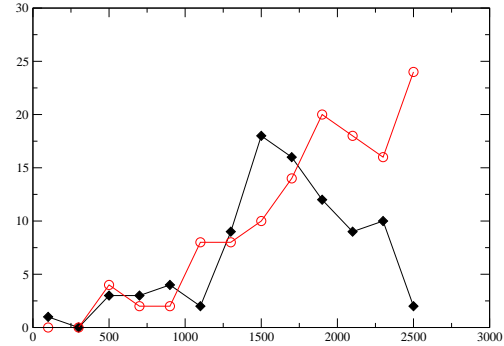}
\caption{ \label{counts} The "arithmetic" quark model meson counts 
(circles) fall short of detected resonances (diamonds) in the mass region 
below $1600 \ MeV$.}
\end{figure}

Let me conduct for you the following simple exercise.
Take the meson counts (in the sense of ``letter counts'', ignore any 
spin or isospin degeneracies) from the last PDG listings \cite{pdg}, 
accept the  ``established'' and  ``confirmed'' resonances as good 
candidates and bin  them, say every $200  \ MeV$. To have a simple 
theoretical benchmark, construct an arithmetic $\overline{q}q$ quark 
model in which the quarks cost as discussed in the previous paragraph 300 
or 500 $MeV$ depending on flavour, each angular excitation 401 $MeV$ and 
each radial excitation $700 \ MeV$. The result is plotted in figure
\ref{counts}. 

From the figure one can derive two observations: that the observed number
of mesons is more than expected below about $1600 \ MeV$, leading us to 
conclude that non conventional mesons have already been detected, and 
second that the meson counts drop dramatically thereafter, even below the 
minimal quark model expectations. Therefore, there is ample room for the 
discovery of new resonances, both conventional and nonconventional, in 
the $2 \ GeV$ range.

\subsection{Exotic mesons}

In the conventional quark model, mesons are made of a pair quark-antiquark
whose quantum numbers are very simply constructed. In the center of mass
frame there is only one orbital angular momentum associated to the 
relative coordinate $\vec{q}$ and two spins $s_q$ and $s_{\overline{q}}$.
Coupling the two spins to give total spin $S$ and then the orbital angular 
momentum $L$ to give a total angular momentum $J$ leads to the 
wavefunction $\langle s_q m_1 s_{\overline{q}} m_2 \arrowvert S 
m_s \rangle \langle S m_s L m_L \arrowvert J m_J \rangle 
Y_L^{m_L}(\vec{q})  \chi_q \chi_{\overline{q}}$. 
Parity reverses the sign of $\vec{q}$ and the spherical harmonic yields a 
phase $(-1)^L$, and since the intrinsic parities of particle and 
antiparticle are opposite, the total parity is $P=(-1)^{L+1}$. 
To form eigenstates of charge conjugation the two particles need to have 
the same flavour, in which case applying $C$ reverses their respective 
role and therefore we collect a phase from the spin Clebsch-Gordan and a 
phase from the spherical harmonic $C=(-1)^{L+S}$.
Giving now integer values to $L$ and $S$, with $J$ taking the values 
$\arrowvert L - S \arrowvert $ to $L+S$ we can construct the ordinary 
meson quantum numbers $J^{PC}$:
$$
0^{-+},1^{--},1^{+-},0^{++},1^{++},2^{++},2^{-+}, ...
$$
From the list will be missing the quantum numbers
$$
0^{--}, \ {\rm even}^{+-}, \ {\rm odd}^{-+} 
$$
and a meson in these channels is called $J^{PC}$-exotic, whereas a 
meson with ordinary quantum numbers but potentially higher Fock space 
content than $\overline{q} q$ is called a cryptoexotic or hidden exotic. 
We can also define flavour exotics to be those mesons with isospin equal 
or higher than $3/2$ or strangeness equal or higher than 2, 
since their minimal $\overline{q} q$ assignment is not allowed. 

\subsection{If it's not forbidden, it's mandatory}

This old saying of quantum mechanics has a two-fold meaning in exotic 
spectroscopy. On one hand, the theory of the strong interactions does not
forbid the existence of exotic mesons. They will therefore be found.
Still they are obviously not the dominant form of strongly interacting 
matter, 
rather an oddity. This can be explained by the fact that they are 
subleading in a large $N_c$ expansion, and by the mass gap that extra 
particles cost. 

On the other hand this makes cryptoexotics a difficult task. All mesons 
with ordinary quantum numbers will have an ordinary $\overline{q}q$ 
leading Fock space assignment to some extent. 
In this respect, one should target searches to mesons with explicitly 
exotic quantum numbers for a clear discovery.  

\subsection{Exotic flavor channels are repulsive}
The benchmark hadronic interactions like $\pi \  \pi$ scattering in an
$I=2$ wave, or $K\ N$ scattering in $S=1$, that had they  shown 
resonant behaviour would
have signaled exotic hadrons, are empirically repulsive. There is a quark 
model explanation within  the Resonating Group Method \cite{ribeiro} which
I briefly sketch. First consider a hadron bound by an attractive color 
exchange modeled with a potential. The single exchange
\begin{equation}
\begin{picture}(150,80)(0,0)
\Oval(10,10)(15,5)(0) \Oval(10,60)(15,5)(0)
\Oval(140,10)(15,5)(0) \Oval(140,60)(15,5)(0)
\ArrowLine(14,70)(136,70) \ArrowLine(16,60)(134,60)
\ArrowLine(14,50)(136,50)
\ArrowLine(14,10)(136,10) \ArrowLine(16,20)(134,20)
\ArrowLine(14,0)(136,0)
\DashArrowLine(60,60)(60,10){3}
\end{picture}
\end{equation}
vanishes because of the color factor (two color singlets cannot exchange a 
color octet). With a Pauli exchange 
\begin{equation}
\begin{picture}(150,80)(0,0)
\Oval(10,10)(15,5)(0) \Oval(10,60)(15,5)(0)
\Oval(140,10)(15,5)(0) \Oval(140,60)(15,5)(0)
\ArrowLine(14,70)(60,70) \ArrowLine(90,70)(136,70)
\ArrowLine(16,60)(134,60) \ArrowLine(14,50)(136,50)
\ArrowLine(14,20)(60,20) \ArrowLine(90,20)(136,20)
\ArrowLine(16,10)(134,10) \ArrowLine(14,0)(136,0)
\ArrowLine(60,70)(75,45) \ArrowLine(75,45)(90,20) 
\ArrowLine(60,20)(75,45) \ArrowLine(75,45)(90,70)
\DashArrowLine(60,60)(60,10){3}
\end{picture}
\end{equation}
the diagram gives a net repulsive contribution. This can happen only
if the wavefunctions of both hadrons overlap (to allow the Pauli exchange) 
and gives rise to the core nucleon-nucleon repulsion that guarantees 
nuclear stability against collapse.

In the case of mesons, the constituent quark model cannot reproduce the 
low energy theorems that guarantee an attractive force to compensate this
repulsion since the constituent quark mass explicitly breaks chiral 
symmetry. A simple field theoretical extension employing the 
Bethe-Salpeter equations \cite{orsay} or the instantaneous Random 
Phase Approximation \cite{cotanchllanes} allows for an annihilation 
diagram
\begin{equation}
\begin{picture}(150,80)(0,0)
\Oval(10,10)(15,5)(0) \Oval(10,60)(15,5)(0)
\Oval(140,10)(15,5)(0) \Oval(140,60)(15,5)(0)
\ArrowLine(14,70)(60,70)  \ArrowLine(14,50)(136,50)
\ArrowLine(16,20)(134,20)
\ArrowLine(136,0)(60,0) \ArrowLine(60,0)(136,70)
\DashArrowLine(60,70)(60,0){3}
\ArrowLine(60,70)(14,0)
\end{picture}
\end{equation}
that provides the necessary attractive force. This diagram 
vanishes unless the flavors of a quark and the antiquark in the other 
meson are equal. Therefore, in exotic flavor scattering where this flavor 
equality is not possible the repulsive interaction is not cancelled and 
exotic flavor channels remain usually repulsive, although with excited 
quantum numbers this can change. In the RPA some 
attractive diagrams in addition to this due to the back-propagating wave 
function are always present and help guarantee the Adler zero, but the 
interaction will remain repulsive although weak. 

The argument is not valid at very low energies (where pion exchange is the 
dominant interaction, the deuteron binds after all) nor higher energies 
where the potential model has little to say.
\section{A glance at the spectrum}

The number of meson resonances is growing rapidly. There are now enough 
established (or firm candidates) to fill up to fourteen $SU(3)$ 
nonets (and meson spectroscopy is starting to resemble plant botanics). 
Some of the assignments below are controversial, but bear them for a 
moment as a starting point for the discussion.
Older than me are the pseudoscalar $0^{-+}: \ \ [\pi K \eta \eta']$ and 
vector $1^{--}: \ \ [\rho \ K^* \ \omega \ \phi]$ nonets that correspond 
to the $L=0$, $S=0,1$ quark model nonets (with a small $D$-wave mixing in the 
vector nonet).
Of course, the charge conjugation assignments given do not
apply to the open flavor mesons, and one should keep in mind 
possible mixing 
between states with equal $J^P$. 

The quark model's $L=1$, $S=0$ nonet is also filled by
$1^{+-}:\ \ [b_1(1235), \ K_{1B},\ h_1(1170,1380)]$.
Late developments \cite{a0f0,scalars} indicate that we have now two scalar 
multiplets, one from the $q\overline{q}$ triplet $L=1$, $S=1$ 
$0^{++}: \ \ [a_0(1450)\  K_0^*(1400)\  f_0(1370,1710)]$ (with all
the $f_0$ mixing, I am just counting states) and what looks like an 
$S$-wave dimeson molecule nonet
$0^{++}: \ \ [a_0(980)\  \kappa(900)\  f_0(600,980)]$.
The pseudovector (flavor) nonet of the (spin) triplet is also filled
with $1^{++}:\ \ [a_1(1260)\ K_{1A}\ f_1(1285,1420)]$ and there are now
two complete tensor nonets, for example the assignments
$2^{++}:\ \ [a_2(1320) \ K_2^*(1430)\ f_2(1430,1525)]$ and
$2^{++}:\ \ [a_2(1700)\ K_2^*(1980)\ f_2({\rm various})]$. 
(or exchange one of the $f_2$ by the $f_2(1270)$).
They lie too 
close to each other for radial excitations, so again we have a likely 
manifestation of meson molecules or four quark states with conventional
$q\overline{q}$ mesons. Radial excitations do appear in the spectrum, and 
already in complete nonets. We can fill two more pseudoscalar nonets
$0^{-+}: \ \ [\pi(1300) \ K(1460)\ \eta(1295,1440_H)]$ and
$0^{-+}: \ \ [\pi(1800) \ K(1830)\ \eta(1760,2225?)]$ and almost two more 
vector nonets  $1^{--}: \ \ [\rho(1450)\ K^*(1410)\ \omega(1420) 
\phi(1680)]$ and $1^{--}: \ \  [\rho(1700)\ K^*(1680)\ \omega(1650)\
\phi({\rm missing})]$. The higher one corresponds mostly to a $D$-wave, 
the lower to the radial excitation \cite{isgur}. 
The lack of a $\phi$ meson at this scale is quite a 
puzzle and a challenge to DA$\Phi$NE. A comment from the audience informs 
us that in the $B$ factories the next $\phi$ seems to appear well above 2 
$GeV$. Even in the likely case of ideal mixing that would imply quite a 
high mass: we expect a value around 1.9 $GeV$. There are a number  of 
other $\rho$ resonances reported, $\rho(1900)$, $\rho(2150)$... that 
having open flavor, should be accompanied by corresponding $\omega$ and
$\phi$ mesons independently of their Fock space assignments. So there are
plenty of opportunities for DA$\Phi$NE to clarify the situation if the 
beam energy is increased to 2 $GeV$.
To complete the discussion let me comment on the higher angular momentum 
multiplets. The $D$-wave $L=2$, $S=0$ also seems to be complete with
$2^{-+}:\ \ [\pi_2(1670)\ K_2(1770)\ \eta_2(1645,1870)]$. The 
corresponding 
$S=1$ (mixing to $G$-wave possible) can be also assigned to
$3^{--}:\ \ [\rho_3(1690)\ K_3^*(1780)\ \omega_3(1670)\ \phi_3(1850)]$. 
Then 
a quark model $F$-wave, $4^{++}:\ \ [a_4(2040)\ K_4^*(2045)\ 
f_4(2050,2300)]$ and unless one is  very interested in Regge theory or how 
the strong interactions depend on angular momentum, candidates to fill 
higher multiplets can be ignored for now.

\subsection{Some mismatches}

The particle tables collect an assortment of extra resonances that I do 
not mention here \cite{pdg}. A few are worth remarking though. There is 
an extra $f_2$ at low energy (I left out the $f_2(1270)$ for a 
cryptoexotic candidate, by which I mean a linear combination of the low 
energy $f_2$), then the $\eta_L(1440)$, both containing to some extent a 
four-quark component. Then an extra $f_0$ where again I left the 
$f_0(1500)$ to represent the appropriate combination of the $f_0$'s that 
likely
construct the glueball state. If the dubious $K_2(1580)$ is confirmed we 
might have a $K_2$ in excess with the $K_2(1770)$ and $K_2(1820)$. 
Then some resonances are obviously missing: a $b_1$ around $1600 \ MeV$ 
would complete a second $1^{+-}$ nonet, an $a_0$ around 1800 would be 
an interesting  addition to fill a $0^{++}$ nonet, there being a reported
$K^{0*}(1950)$ (this automatically would pull two of the $f_0$'s out of 
the glueball candidate list, and there are some reported in the $2\ GeV$ 
region). Also a $K_1$ in the $1.8\ GeV$ region would make some $J=1$ 
mesons fall 
in place, and outstandingly for this conference's purpose, a $\phi$ in 
the $1.9\ GeV$ range is definitely expected.

\subsection{And some exotica}

The firm candidates for explicitly exotic mesons are the $1^{-+}$ broad
structures with mass and width $M=1380(20) \ MeV$, $\Gamma=300(40) \ MeV$
and $M=1600(25)\ MeV$, $\Gamma=310(60)\ MeV$. The first has a decay
mode into $\eta \pi$, the second into $\eta' \pi$. A sensible account of
the current status (and my favorite interpretation, see below) of these
resonances is given in \cite{adamsletter}. There might now be a
third candidate around $1.9\ GeV$ \cite{meyer}.
For very long these two states have been accused of being hybrid mesons,
specially the second since it decays to the supposedly ``glue rich''
channel with an $\eta'$. These arguments do not resist closer examination
and the trophy for a hybrid meson is still open.
There is also a very interesting candidate, $X(1600)$ with reported 
isospin 2. If confirmed, this would be the first case of a meson 
with a {\it guaranteed} four-quark leading wavefunction \cite{filippi}.

\subsection{Identification by decay patterns}

Beyond mass and width assignments, a detailed understanding of mesons 
requires predictions for their branching ratios into different possible 
open channels. The favorite model for these calculations is the $^3P_0$ 
model (T. Barnes has just completed a short historical account where 
references can be tracked \cite{barnes}). In this mechanism, completing 
quantum-mechanical models of mesons such as the Constituent 
Quark Model or the Flux Tube Model at a fixed particle number, a 
$q\overline{q}$ pair is pulled out of the Fermi sea, and a rearrangement 
of color by a Pauli exchange leads to a two-meson decay. For example, for 
a conventional meson one would have:
\begin{equation}
\begin{picture}(150,80)(0,0)
\Oval(10,10)(15,5)(0) 
\Oval(140,10)(15,5)(0) \Oval(140,60)(15,5)(0)
\ArrowLine(14,20)(136,20)
\ArrowLine(136,0)(44,50) \ArrowLine(44,50)(136,70)
\BCirc(44,50){3}
\ArrowLine(136,50)(14,0)
\end{picture}
\end{equation}
Within the $^3P_0$ model there are extensive decay calculations assisting 
experimental searches of hybrid mesons \cite{page} in the flux tube model. 
(The decays of ordinary mesons have also been extensively studied). An 
outstanding prediction for hybrid mesons is their preference to decay to a 
pair of $S$-$P$ mesons.

Let us also adopt the point of view of a Quantum Field Theory Hamiltonian. 
The first task is to perform a diagonalization in the Fock space to find 
the representation of the mesons in each channel, say for exotic meson 
$X=\alpha \arrowvert q \overline{q} g\rangle+\beta \arrowvert 
q\overline{q}q\overline{q}\rangle + ...$.
Then, once the Hamiltonian has been exactly diagonalized, the leading
decay to two mesons proceeds by the wavefunction overlap of the four-quark 
component of the state with the state of two mesons streaming freely to 
the detector. That is, all two meson decays proceed via ``Fall-Apart'' 
decays of the four quark component. For a hybrid this mechanism can be 
depicted as
\begin{equation} \label{fallapart}
\begin{picture}(150,150)(0,0)
\put(0,135){$\alpha$}
\Oval(20,135)(15,5)(0)
\ArrowLine(24,145)(75,145)
\ArrowLine(24,125)(75,125)
\Gluon(26,135)(75,135){3}{6}
\put(15,95){$+$} \put(0,30){$\beta$}
\Oval(20,30)(30,8)(0)
\Oval(140,10)(15,5)(0) \Oval(140,50)(15,5)(0)
\ArrowLine(26,20)(136,20) \ArrowLine(20,60)(136,60)
\ArrowLine(136,0)(81,20)  \ArrowLine(81,20)(26,40) 
\ArrowLine(136,40)(78,20)  \ArrowLine(78,20)(20,0)
\end{picture}
\end{equation}
where any possible rescattering between the fermion lines in the second 
diagram should have already been taken into account in the construction of 
the various meson-fermion vertices, and the annihilation or absorption of 
the gluon in the first diagram should likewise already have been taken 
into account in the diagonalization of the Hamiltonian to construct the 
total wavefunction of the state.
This diagram shows that {\it all} exotica decay to lowest order via its 
minimal multiquark component in a field-theoretical description (where 
``lowest order'' refers to the wavefunctions of the final state mesons, 
that taken as conventional mesons, start with $q\overline{q}$, that is, 
lowest order in the constituent mass gap). 
Therefore, the hybrid wavefunction component of a meson hides behind the 
four (six) quark component as hadronic decays are concerned. Since the 
gluons do not carry electric charge, similar considerations apply to 
radiative decays. 
This shows that until we have an excellent grasp of the physics of 
four and more quark states, we will not be able to fully trust predictions 
on the possible decays of exotica. 

The result is unlike a lottery, where purchasing a ticket (detecting 
a $J^{PC}$-exotic meson) gives you a winning chance (an explicit gluonic 
excitation in the spectrum). Here until you have all the tickets in your 
hand you don't collect your prize, since unravelling the wavefunctions
in (\ref{fallapart}) is an arduous theoretical task, and we require 
knowledge of all excitations in the reasonable energy range for a given 
channel.

\section{Explicit glue} \label{gluons}

\subsection{Glueballs}

\begin{figure}[htb]
\centering
\includegraphics*[width=75mm]{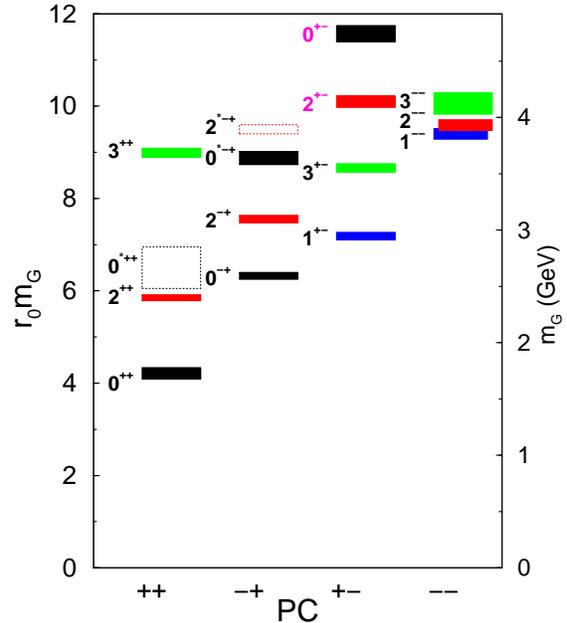}
\caption{ \label{glueballspec} The lattice glueball spectrum (C. 
Morningstar and M. Peardon).}
\end{figure}

Glueballs are an interesting theoretical construction. Even without 
quarks, since chromodynamics is a non-abelian theory with non-linear gluon 
self-couplings, there would still be mesons. The spectrum has been 
calculated in the lattice on a number of occasions. The result of a 
well-known calculation is reproduced (courtesy of the authors of 
\cite{morningstar}) 
in figure \ref{glueballspec}. 

One sometimes wonders how inputing random numbers, valuable organized
numbers are output, but this is so:
the salient features of this spectrum are very well captured in 
model terms \cite{sscj}. If one attempts to couple two massive 
constituent vector bosons in an S-wave to form a bound state, the possible 
quantum numbers would be $J^{PC}=(0,1,2)^{++}$. By observing figure 
\ref{glueballspec} we see that the $1^{++}$ state is absent from the 
low-lying spectrum. Therefore a massive constituent gluon model fails. 
But a model with transverse gluons, such as based in the Coulomb gauge,
automatically succeeds thanks to Yang's two-photon theorem. In these 
models, a gluon BCS mass-gap equation is solved that generates a gluon 
mass dynamically. Then the gluons can maintain their transverse nature since 
there is no explicit mass term and, being bosons, cannot couple to spin 1.
Solution of the Tamm-Dancoff equation for the two-body problem provides a 
spectrum similar to the lattice results at low energy. 

Another obvious feature in figure \ref{glueballspec} is that odd-parity 
glueballs are heavier. This in two-gluon models follows from the necessity 
of a p-wave. The wavefunction $\langle s_1 m_1 s_2 m_2\arrowvert S m_s 
\rangle Y_L^{m_L}$ predicts the $P$-wave glueballs to have 
$J^{PC}=(0,2,3)^{-+}$. 
Finally, the negative charge conjugation states 
are even heavier, also natural in a model where a third gluon would be 
necessary (and again the mass-gap lifts this state).

The gluon mass gap can be predicted from this lattice data to be about 
800-900 $MeV$. The lightest scalar glueball thus appears at 1600 $MeV$ (or 
above) and the tensor glueball separated by a hyperfine splitting, 
slightly above 2 $GeV$. This mass gap is tied to the string tension 
calculated in the same lattice (the same happens in model calculations). 

Finally the obvious remark that gluons carry no flavor quantum numbers, 
and therefore glueballs appear only as singlets in the spectrum, stirred 
through the $f_0$, $f_2$ families. A study to disentangle the glueball 
components \cite{closekirk} (with the approximation of ignoring four quark 
states) has found the scalar glueball to be  shared between 
the $f_0(1370)$ and $f_0(1500)$ with some component in the $f_0(1710)$. 
This is based on decay predictions from flavor $SU(3)$ symmetry for 
$u\overline{u}$, $d\overline{d}$ and $s\overline{s}$ pairs and gives a 
qualitative picture of how difficult cryptoexotica may be: the only 
manifestation of this glueball state seems to be a supernumerary scalar 
meson.

\subsection{Glueballs fall on Regge trajectories}

\begin{figure}[htb]
\centering
\includegraphics*[width=65mm]{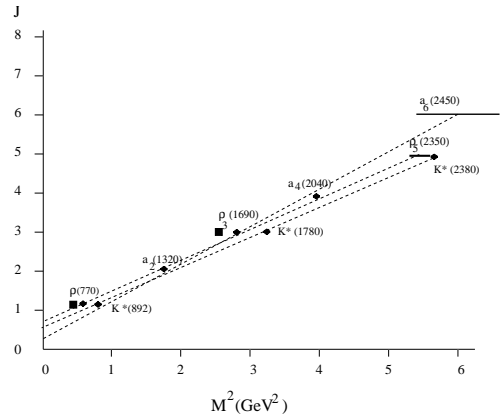}
\caption{ \label{reggemesons} Some meson Regge trajectories.
}
\end{figure}

Short of direct detection, a second window to glueballs and the gluon mass 
gap is provided by high energy physics. The interaction between static 
color sources at long distance follows a linear behavior, according to 
lattice QCD studies. This provides the theoretical basis for the linear 
potential and for the observation that mesons fall on straight lines in a 
plot of $J$ versus $M^2$ (see figure \ref{reggemesons}).

Similar plots can be produced for baryons lending support to a diquark 
wavefunction clustering, but that is a theme for a different conference.

In Coulomb gauge QCD the potential interaction acts between color charge
densities, and therefore the color charges associated to the gluons (after 
dynamical chiral symmetry breaking, also static charges) interact with a 
similar potential. A difference is the color factor for the potential 
exchange, 3 for gluon-gluon as opposed to 4/3 for $\overline{q}q$. 
This changes the slope of the Regge trajectories and brings them close to 
parallel to the famous pomeron Regge trajectory
\begin{equation}
J=1.08 + 0.25 t
\end{equation}
equating $t=M^2$ we see that glueball exchange probably plays a role in 
total cross sections at high $s$, low $t$.

\begin{figure}[htb]
\centering
\includegraphics*[width=75mm]{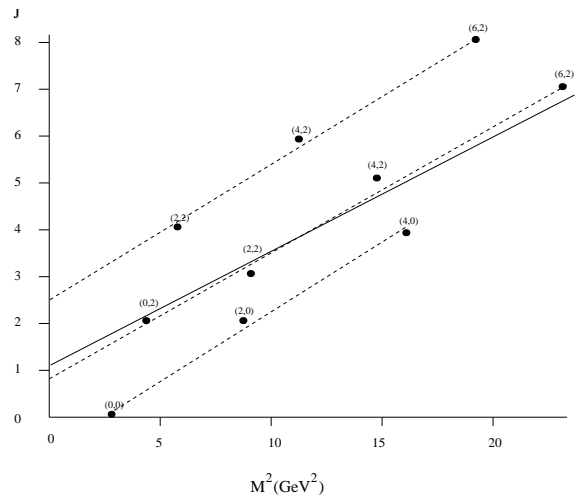}
\caption{ \label{glueballregge} BCS/Tamm-Dancoff glueball spectrum with
timelike Cornell potential between color charge densities. The 
wavefunctions are constructed in the Russell-Saunders scheme and 
configuration mixing is  not included. The solid line
is the pomeron Regge trajectory. States following to the left of it are
unphysical and their low mass is probably due to excessive spin-orbit
coupling.}
\end{figure}
\begin{figure}[htb]
\centering
\includegraphics*[width=75mm]{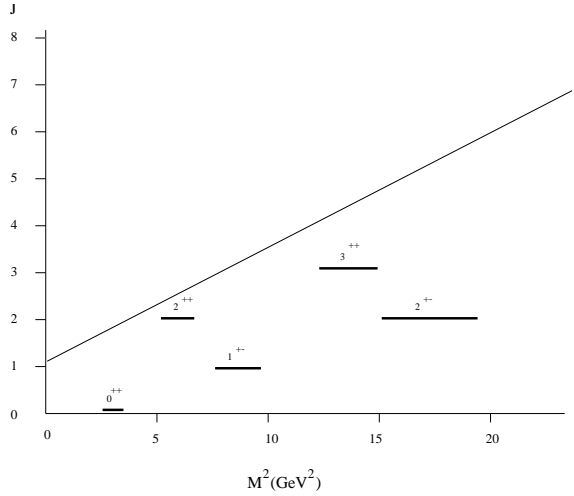}
\caption{ \label{latticeglueball} Lattice glueball states also seem to
fall on Regge trajectories. The $4^{++}$ state, with larger error bars,
recently calculated, is compatible with the pomeron trajectory.}
\end{figure}

Our glueball calculations \cite{sscj,todosglue}, based on QCD time-like 
vector exchange, suffer 
from an excessive spin-orbit coupling. This is analogous to the same 
phenomenon in the quarkonium spectrum, that leads spectroscopists to 
hypothesize scalar confinement. Therefore, to extract relatively 
model-independent information from our model glueball spectrum, we should 
look at states with a vanishing expectation value of $L\cdot S$. This is 
achieved if either $L=0$ or $S=0$. For S=0 there is a Regge trajectory 
formed by the glueballs $L=0,\ 2,\ 4,...$, $J^{PC}=(0,\ 2,\ 4...)^{++}$ 
that passes by the lightest scalar glueball, parallel and below the 
pomeron trajectory. From this, the slope of all glueball Regge 
trajectories can be extracted. 
There is still the $L=0$ case. With $S=0$ we recover the scalar glueball. 
But with $S=2$, the lowest-lying tensor glueball $2^{++}$ does not suffer 
from spin-orbit corrections and therefore we can confidently calculate its 
mass. We find it just above $2\ GeV$, and almost within lattice error 
bands, it falls right on the pomeron Regge trajectory. Other glueball 
states with quantum numbers $J^{++}$ calculated in the lattice seem to 
also fall on the pomeron trajectory, here the error bands being 
considerably larger.
 
In a more elaborate model we would expect glueballs with $L=2$, $L=4$, 
... and spins aligned to also fall on or near the pomeron Regge 
trajectory. Currently this is not the case since our large spin-orbit 
coupling pushes these states left of the pomeron trajectory (in
contradiction with total energy cross section growths). Therefore further 
work is required to improve the current models. Similar results are 
reported by other authors \cite{adamlosalamos} with Regge trajectories 
that start bending and behaving non-linearly.

As a conclusion, we expect fits to high energy cross-sections to provide a 
complementary approach to low energy spectroscopy in the understanding of 
gluonic excitations.

It is also interesting that the odderon, Regge pole exchange with negative 
$C$-parity, would produce differences between the $p\overline{p}$ and $pp$ 
cross sections that are not experimentally observed. Within our 
constituent approach this is again easy to understand, as negative parity 
glueballs require a third gluon and thus lie to the right of the pomeron 
trajectory and are less relevant for high $s$ scattering.

\subsection{Hybrid Mesons: quantum numbers}

Hybrid mesons are also defined only with the help of a model. 
From the few-body point of view \cite{hybrids,kalas} in that the 
quark-antiquark pair are accompanied by an explicit constituent gluon
(or in our models, a quasiparticle-gluon after dynamical chiral symmetry 
breaking), it is a relatively simple task to predict the quantum 
numbers of lowest lying hybrid states. Consider the three-body system 
in its center of momentum (CM) frame. The remaining six momentum space 
coordinates can be chosen as 
\begin{equation}
q_+=(q+\overline{q})/2\ \ \ q_-=q-\overline{q}
\end{equation}
(the gluon's momentum is immediately determined in the CM frame).
There are three particle spins, the quark and antiquark $s_q$ and 
$s_{\overline{q}}$ and the gluon spin $s_g$. All 
angular momenta can be then added 
\begin{equation}
s_q + s_{\overline{q}} + s_g +L_+ + L_- =J
\end{equation}
with the usual angular momentum rules. Of the several possible recouplings 
we choose to combine $s_q$ and $s_{\overline{q}}$ to spin $S$ since then 
one can form (for equal $q\overline{q}$ flavor) eigenstates of charge 
conjugation with eigenvalue
\begin{equation}
C=(-1)^{1+S+L_-} \ .
\end{equation}
Since the gluon is a vector state it brings an additional $(-1)$ to the 
parity computation yielding
\begin{equation}
P=(-1)^{L_++L_-} \ .
\end{equation}
The lowest lying states are expected to be those with $L_+=L_-=0$, and the 
possible combinations are therefore $J^{PC}=1^{+-},\ (0,1,2)^{++}$. These 
have conventional quantum numbers and mix with ordinary $q\overline{q}$ 
mesons. Allowing for a $P$ wave we have either
$L_-=1$, $J^{PC}=0^{-+},\ ({\mathbf 1},2)^{-+},\  ({\mathbf 
0},1,2,3)^{--}$ or $L_+=1$, $J^{PC}=(0,{\mathbf 1},2,{\mathbf 
3})^{-+},(1,2)^{--}$ with various multiplicities due to the intermediate 
spin states. 

In our model calculations the $L_+$ excitation is less expensive 
energetically ($L_-$ can be taken as the variable conjugate to the 
quark-antiquark position, that see a net repulsion in a color octet)
Therefore the lightest exotic hybrid mesons have 
quantum numbers $1^{-+}$, $3^{-+}$ and $0^{--}$ in this approach.

\subsection{Hybrid mesons: masses}

Our model calculations with a variational approach and a modest 
wavefunction basis (hence, upper bound to the minimum eigenvalues) 
concludes there are no hybrid states, below $2\ GeV$. This is somewhat 
high compared to lattice expectations that estimate the first hybrid to 
lie around $1.9\ GeV$. In this there is not unanimous agreement, as 
recent calculations point to $1.7\ GeV$ \cite{milc} and $2.1\ GeV$ 
\cite{chinos} for the lowest lying $1^{-+}$ exotic hybrid.

As flavor is concerned, hybrids come in flavor nonets as their regular
$q\overline{q}$ counterparts. Isospin $I=1$ hybrid mesons are expected to 
be lighter than $I=0$ as the following annihilation diagram

\begin{equation}
\begin{picture}(100,50)(0,0)
\Oval(20,25)(20,6)(0) \Oval(80,25)(20,6)(0)
\ArrowLine(22,45)(50,25) \ArrowLine(50,25)(28,25)
\ArrowLine(50,45)(78,45) \ArrowLine(72,25)(50,45)
\DashArrowLine(50,25)(50,45){3}
\Gluon(22,5)(78,5){3}{6}
\end{picture}
\end{equation}
is only possible for isospin zero, and adds about $200\ MeV$ to the mass 
of these states.

An alternative description is provided by the flux tube model 
\cite{fluxtube}, inspired in lattice QCD in the limit of strong coupling. 
Not surprisingly, the model (usually combined with the Born-Oppenheimer 
approximation for heavy quarks) is in good agreement with lattice data for
$bg\overline{b}$ states \cite{juge}. 
For light hybrid states, the model predicts isospin 1 multiplets with 
quantum numbers $({\mathbf 0},1,{\mathbf 2})^{+-}$, $(0,{\mathbf 
1},2)^{-+}$, $1^{++}$, $1^{--}$ at about $1.9 \ GeV$.
The mass difference with our approach can be explained by the fact that 
this string simulating a flux tube provides for a color singlet potential 
(therefore attractive) 
between the quark and antiquark also in the excited (hybrid) 
configuration, whereas in an approach with an explicit QCD gluon the quark 
and antiquark are in a color octet configuration, repulsive. 
If the original flux-tube model is corrected employing the lattice excited 
adiabatic potential instead of the attractive Coulomb tail from Isgur and 
Paton, the mass predictions rise again to above $2 \ GeV$ \cite{katja}.

In any case, both approaches concur to predict exotic hybrids to be well 
above the two experimental candidates (in disagreement with old bag model 
calculations \cite{baghybrids}). 

Given the large number of channels and possible angular momentum 
combinations, the spin splittings due to fine and hyperfine interactions 
in hybrid mesons, that lift the degeneracies between the various $J^{PC}$, 
are quite 
intricate. Within our few-body approach, the 
relativistic structure of the Hamiltonian provides for some splittings, 
but the $\gamma_0$ time-like vector potential is known to be deficient in 
the ordinary meson sector, leaving this as an open issue. 
 These splittings have been calculated in the context of the
Flux Tube Model  \cite{katja} and lift the degeneracy between the 
vector $1^{--}$ and exotic $1^{-+}$ hybrid mesons.

All theoretical approaches seem to concur that vector hybrids in the 
charmonium system should appear at about $4.4 \ GeV$. In our calculations, 
up to four hybrid states appear around the last known $\psi(4415)$ 
resonance (and suggest above it a continuum that would require careful 
work to discern the various states).
 \section{Multiquark states}
\subsection{Four quark states}
Constructing the angular momentum wavefunctions for 
$qq\overline{q}\overline{q}$ is long to 
describe, so let me do it with a picture:
\begin{picture}(200,200)(0,0)
\ArrowLine(10,170)(50,170)
\ArrowLine(10,150)(50,150)
\ArrowLine(30,170)(30,150)
\BCirc(10,170){8}  \BCirc(10,150){8} 
\BCirc(50,170){8}\BCirc(50,150){8} 
\put(8,168){1} \put(8,148){$\overline{2}$} 
\put(48,168){3} \put(48,148){$\overline{4}$}
\put(90,170){$s_1$} \put(90,155){$s_3$}\put(90,140){$L_{13}$}
\ArrowLine(100,172)(115,162) \ArrowLine(100,157)(115,162)
\put(115,160){$S_{dq}$}
\ArrowLine(130,162)(140,152) \ArrowLine(105,142)(140,152)
\put(140,150){$j_{dq}$}
\put(90,110){$s_2$} \put(90,95){$s_4$} \put(90,80){$L_{24}$}
\ArrowLine(100,112)(115,102) \ArrowLine(100,97)(115,102)
\put(115,100){$S_{d\overline{q}}$}
\ArrowLine(130,102)(140,92) \ArrowLine(105,82)(140,92) 
\put(140,90){$j_{d\overline{q}}$}
\ArrowLine(155,90)(165,122) \ArrowLine(155,150)(165,122) 
\put(165,120){$S$}
\put(90,50){$L_{12-34}$}
\ArrowLine(175,122)(190,77) \ArrowLine(115,55)(190,77)
\put(190,75){$J$}
\put(0,10){$P=(-1)^{L_{13}+L_{24}+L_{12-34}}$}
\put(0,30){$C=(-1)^{S+L_{12-34}}$}
\end{picture}

With all $L=0$, the wavefunctions that can be constructed have 
conventional quantum numbers $(0,2)^{++},\  1^{+-},\  1^+$.
If we allow  a $P$-wave, then also possible are $1^{--},\ (0,1,2,3)^-,\ 
(0,{\mathbf 1},2)^{-+}$. These assignments can proceed in various ways 
through the angular momentum tree, and therefore there are several 
possible constructions of each state. This leads to a rich spectrum first 
calculated in the bag model \cite{jaffe} that leads to the so called {\it 
state inflation} not experimentally observed, although the light scalars 
at least seem to fill the $S$-wave nonet. They are very broad as expected 
for wavefunctions with an OZI superallowed ``fall apart'' decay.

Notice the (unfortunate?) coincidence that the lowest exotic seems to be 
the $1^{-+}$ and that four quark states are predicted to be lighter than 
hybrids. With the angular momentum construction above there is a unique 
way of constructing this state. But to build an isospin 1 state one can of 
course resort to $(u\overline{u}+d\overline{d})u\overline{d}$ or
$s\overline{s}u\overline{d}$, and therefore two light, broad four quark 
structures are expected in agreement with observation. In other channels 
they may just be assigned to background as more prominent $q\overline{q}$ 
resonances appear, but in this exotic channel they have to be the first 
structures appearing.  In this sense, the 
other exotic quantum numbers that hybrid mesons span are more promising.

On occasion of another DA$\Phi$NE workshop, Badalyan presented his results 
for four quark states \cite{badalyan}. Since these calculations have to my 
knowledge not been superseded by subsequent studies (given the lack of 
experimental motivation), I abstract in table \ref{4q} the 
result for the center of gravity of the spin multiplets for the ground 
state. In both Jaffe's and Badalyan's work the scalar mesons are lightest, 
and the large (unreliable) spin splittings make further progress 
difficult. The lightest exotic $1^{-+}$ was predicted at $1.7\ GeV$.

\begin{table}[hbt]
\begin{center}
\caption{\label{4q}
In this table we give the center of mass (prior to spin splittings) of the 
four quark ground state 
in terms of its flavor content. Ideal mixing is expected to hold in 
four-quark systems. \vspace{0.3cm}} 
\begin{tabular}{c|cc}
\hline
M(flavor) & Badalyan(1987,1991) & Jaffe(1977) \\
\hline \\
4 light q & 1565 & 1540 \\
2 light 2 s & 1950 & 1800 \\
4 s & 2260 & 2150
\end{tabular}
\end{center}
\end{table} 
Another problem with these calculations is their lack of agreement with 
the low energy pion theorems: in channels where broad structures decay to 
light Goldstone bosons, one would expect chiral symmetry to play a major 
role. Only recently it was understood how to incorporate chiral symmetry 
into calculations beyond the spectrum through the RPA/Bethe-Salpeter 
approach \cite{ribeiro2} and we may expect progress in this 
direction.

In the vector $1^{--}$ channel the four quark {\it state inflation} leads 
to a prediction \cite{badalyan,mulders} of a large number of states, at 
1500, 1660, 1830, 1860, 1940, 2000, 2070 $MeV$ and this is not acceptable with 
our current understanding of the data. Here is interesting work for 
theorists. Initially we may assume they are broad and overlaping.

An exception to the rule that four-quark states are broad can 
be found in the $f_0(980)$ because it is just below its natural two kaon 
decay. This may happen again to some extent, so a number of interesting 
thresholds (like the two $\omega$, two $\phi$ or two nucleon) are worth 
detailed scrutiny.

Among theorists, there has been some skepticism about four-quark states 
(with the exception of the light scalars), that can (other than by 
lack of direct evidence) partly be tracked to 
the work of Weinstein and Isgur \cite{weinstein} who, conducting a 
variational calculation containing $qq\overline{q}\overline{q}$ in the 
basis, found separate hadron states to be lighter than 
compact multiquarks. This has the advantage of providing some support for 
the nuclear physics picture where nuclei are clusters of nucleons, not of 
quarks.
 
\subsection{The Pentaquark}

This expectations are at odds with the recent detection \cite{nakano} 
of a pentaquark. Although a baryon, this state opens new perspectives in 
meson spectroscopy. This state, at $1540 \ MeV$, has been observed to 
decay 
to $pK_0$ and $nK^+$. The threshold for this channel is $1435 \ MeV$ and 
therefore there is sufficient phase space for the decay. The width of this 
$\Theta^+(1540)$ state is intriguingly narrow, less than 10 MeV. More 
interesting is the fact that its leading wavefunction assignment in Fock 
space has to be $\arrowvert uudd\overline{s}\rangle$ and is therefore 
flavor exotic. If further confirmed beyond the present experimental data, 
this state supposes a true revolution in spectroscopy. 

We could ask ourselves if this state is a molecule in the sense of the 
$f_0(980)$ or deuterium. But the $KN$ interaction is repulsive in an 
$S$-wave, and only very mildly attractive in a $P$-wave (recall our 
general discussion about exotic channels being repulsive). Therefore the 
system does not resonate. It has been suggested \cite{marques} that a pion 
would stabilize the system to form a three body state $K\pi N$. We have 
performed standard calculations within the Chiral Lagrangian 
supplemented with unitary (Lippman-Schwinger) techniques \cite{oset} 
and found there is indeed attraction in the $J^P=\frac{1}{2}^+$ 
channel preferred by theorists, but way insufficient to bind the 
system. This is consistent with the low energy database accumulated 
through the years: only at higher energies around $1800\ MeV$ there have
been persisting hints \cite{colombia} of resonances that could fit into a 
molecular-type approach, collectively called $Z^*$. 

Therefore  if this state is confirmed, it will have to be assigned to a 
compact pentaquark structure bound by QCD forces. And open many questions 
as to what other multiquark states are there, and where.

\subsection{Six quark states}
At least one six quark state obviously exists (deuterium). 
It can be argued that it is totally a molecular state given its
large radius and small binding energy.  The question
here is whether a six quark meson can be found. The obvious
place to look is just under (around?) the two nucleon threshold.
Along the years intermitent data have supported the existence
of baryonium, with no conclusive evidence to date. 

Lately there are promising candidates in the reaction
$J/\Psi\to \gamma \ p\overline{p}$ at BES \cite{bes} and in multipion 
production processes \cite{deFalco}.
The BES results are compatible with a $0^{-+}$ resonance, that would  
suggest parabaryonium has been found (the analogous of the pion with 
$L=0$, $S=0$, instead of a quark, a proton, and instead of an antiquark, 
an 
antiproton). This state has the proton and antiproton spins antialigned, 
and it should be accompanied by an almost degenerate vector state, 
$1^{--}$ (orthobaryonium) that is a prime candidate for searches at 
DA$\Phi$NE with an enhanced energy beam. 
These resonances are expected to be narrow (by the necessity of 
annihilating that large number of valence quarks and the OZI rule), just 
below $NN$ threshold and with suppressed $K\overline{K}$ decays (no 
strange valence content in either of the constituents). 

An interesting question is whether this state appears in all possible 
channels $n\overline{n}$, $n\overline{p}$, $p\overline{n}$, 
$p\overline{p}$, or just in the latter. In this case we could still argue 
on a nucleon-antinucleon state loosely bound by QED forces alone,
violating isospin symmetry (alternatively something totally different as 
one of the mentioned four-quark states or a hybrid state would be a 
possibility in this mass range).

In any case, since the quark model assignments for vector $q\overline{q}$ 
states predict the $D$-wave vector comfortably close to the $\rho(1700)$ 
and the next radial excitation is above $2\ GeV$, any vector resonance in 
the interval of energies $1.8-2.0\ GeV$ is a prime cryptoexotic candidate.

\section{Conclusions and Outlook}

I hope to have conveyed to you some of the excitement in light exotic 
spectroscopy. In my view, below $1.6 \ GeV$ we have already a state 
saturation and the identification of the wave function components of four 
quark states is the most interesting problem. 
Above $1.6\ GeV$ many conventional, let alone exotic and cryptoexotic are 
missing. To what extent they can be separated from one another and the 
continuum they form in most hadronic processes is a question of 
intelligent filters, multiple particle decay analysis and hard labor.

The most intriguing excitations are those containing some sort of glue -- 
glueballs and hybrid mesons, but they are hiding behind the purely quark 
(let it be two or four...) wavefunctions in each channel. The model is the 
scalar glueball, probably already in our pocket as a supernumerary $f_0$ 
(in fact, as a good quantum state, in both pockets at the same time). We 
expect the situation with the tensor glueball to be the same: just hard 
work ahead in the $2^{++}$ meson channel.

For hybrid mesons the situation would seem cleaner because of the 
possibility of exotic quantum numbers. But this is misleading since four 
quark states hide a hybrid as effectively as two quark states hide a 
glueball. The experimentally studied case, with $1^{-+}$ quantum numbers 
is the model. Other exotic waves should (and probably will, in experiments 
like COMPASS or GLUE-X) be explored.

To correctly identify these excitations they need to be separated from 
four-quark states, about whom very little is known. In turn, the best 
grasp on these states comes from flavor exotic waves. These are generally 
repulsive at low energies, but the possible detection of a pentaquark now 
should make us rethink about them. At higher energies more careful scans 
should be conducted.

What contribution can DA$\Phi$NE make? If the other important physics 
issues addressable by the upgraded machine allow the beam energy to be 
increased to 2 $GeV$, then with its high luminosity it could make the best
map of the $1.7-2\ GeV$ energy interval where some interesting states may 
appear. To be really competitive with other experiments though, $2.5\ GeV$ 
would be much better. Then the vector isoscalar channel could be really 
mapped out to much higher energies, and the decay products would naturally
span the $1.5-2\ GeV$ region with different quantum numbers.

The interested reader can find more information in the recent overview of 
Curtis Meyer \cite{meyer} where lots of tables and data are given that I 
do not reproduce here. Another source of information is the Exotica Web 
page http://fafnir.phyast.pitt.edu/exotica/  .

\subsection{Acknowledgments}

First I want to thank the present and former group at North Carolina State 
University 
for attracting my interest to the problems of exotic 
spectroscopy. I also would like to thank the organizers of this workshop 
for the invitation to give a perspective of the field. I tried to keep the 
presentation pedagogic to attract the wider audience, and apologize 
to a few experts attending the session who had to hear again known 
statements. I also would like to apologize to the authors of an inmense 
number of papers that compose our knowledge basis of the field and whose 
names do not appear in the following minimal reference list.
This work was supported by Spanish government grants FPA 2000-0956, BFM 
2002-01003.

{\emph{Since the first posting of this work two  related 
papers have appeared. One by J. E. Ribeiro 
refreshing on the RGM methods \cite{ribeirolast}, one by E. Swanson, 
whom I thank for some useful comments on the manuscript, 
proposing a four quark state assignment for the new charmonium state 
X(3872) with a specific dynamical model \cite{swansonlast}.}}

\end{document}